\documentclass[twocolumn,runningheads]{svjour2}
\smartqed  
\usepackage{graphicx}
\newcommand{\gtrsim}
	{\mathrel{\raisebox{0.4ex}{\hbox{$>$}}\kern-0.75em
	\raisebox{-0.5ex}{\hbox{$\sim$}}}}
\newcommand{\lesssim}
	{\mathrel{\raisebox{0.4ex}{\hbox{$<$}}\kern-0.75em
	\raisebox{-0.5ex}{\hbox{$\sim$}}}}

\journalname{Astrophysics and Space Science}
\begin{document}

\title{Dark Matter: the Connection with Gamma-Ray Astrophysics}
\subtitle{Discovering Dark Matter particles with astrophysical observations}


\author{Gianfranco Bertone}


\institute{INFN, Sezione di Padova, Via Marzolo 8, Padova I-35131, Italy}

\date{Received: date / Accepted: date}

\maketitle

\begin{abstract}
We review the status of indirect Dark Matter searches, focusing in 
particular on the connection with gamma-ray Astrophysics, and on the
prospects for detection with the upcoming space telescope 
GLAST and Air Cherenkov Telescopes such as CANGAROO, HESS, MAGIC and 
VERITAS. After a brief 
introduction where we review
the fundamental motivations for indirect searches, we tackle the question
of whether it is possible to obtain strong enough evidence
from astrophysical observations, to claim discovery of Dark Matter particles. 
To this purpose, we discuss some recent conflicting
claims that have generated some confusion in the field, and present
new strategies that may provide the long-awaited smoking-gun for Dark
Matter.  
\end{abstract}

\section{Introduction}
\label{intro}

Dark Matter (DM) is one of the pillars of Modern Cosmology, and 
cosmological observations provide the most robust (if not the only)
{\it evidence} for physics beyond the Standard Model. The discovery of Dark Matter
particles may thus represent at the same time the discovery of 
new physics, and the identification of one of the most fundamental 
consituents of the Universe (for reviews see e.g. 
Refs.~\cite{Bergstrom:2000pn,Bertone:2004pz}. 
Of course the best
case scenario is that DM particles are found at accelerators,
in which case astrophysical observations can be used to prove that
the discovered particles have the appropriate cosmological abundance,
and to further constrain their properties~\cite{Hooper:2006wv}.

But what if DM particles are {\it not} observed at accelerators, what
if it takes many years to obtain conclusive answers from, say, 
the analysis of events at the Large Hadron Collider (LHC)? Is it 
possible to obtain conclusive
enough evidence from astrophysical observations {\it only}, 
to actually claim 
``discovery'' of DM particles?

DM searches can be broadly divided in three different categories. 
First, one could search for new particles at accelerators. Much
attention has in fact been devoted to the prospects for detecting
new physics, and in particular signals of Supersymmetry (SUSY) 
with accelerators, such as Tevatron and the LHC.
If new particles are discovered, it might be even possible, starting from the
mass spectrum and particle properties of the new theory, to obtain a 
tentative estimate of the relic density of the new particles, thus 
allowing a tentative identification of the DM particle
(see e.g. Refs.~\cite{Tovey:2002jc,Baltz:2006fm} and references therein).

The prospects for discovery at accelerators have been worked out 
also for an alternative candidate, i.e. the lightest 
Kaluza--Klein particle (LKP)~\cite{Servant:2002aq} , in theories with
Universal Extra Dimensions (UED)~\cite{Appelquist:2000nn}. 
It might be possible already at LHC to discriminate between SUSY
and UED, thus to obtain a first hint on the nature of DM~\cite{Datta:2005zs}.   
Alternatively, one might search for DM particles {\it directly}, i.e.
my measuring the energy recoil of a nucleus hit by a DM particle
streaming through the Earth. More than 20 direct DM detection 
experiments are either now operating or are currently in development.  
We refer the interested reader to Refs.~\cite{Bergstrom:2000pn,Bertone:2004pz}
for more details. 

Finally, DM can be searched for {\it indirectly}, i.e. through the
detection of its annihilation products such as gamma-rays, neutrinos
and anti-matter (e.g. Ref.~\cite{Bertone:2004pz} and references therein). 
Although indirect searches are inevitably affected by
large astrophysical uncertainties, it is nonetheless possible to obtain 
useful constraints on the properties of DM, and possibly, as we argue below, to
obtain strong enough evidence to claim discovery. 

The paper is organized as follows: we tackle some fundamental
questions about the motivations for indirect DM searches in 
Sec.~\ref{sec:faq}, whose title could be  ``Everything you always wanted to know about DM and 
never dared to ask''. The experienced reader can skip this section
and move directly to Sec.~\ref{sec:claims}, where we provide a
critical discussion of some recent conflicting claims of ``discovery''
of DM, while Sec.~\ref{sec:smoking} is devoted to the most recently
proposed strategies that could eventually provide the smoking-gun
for DM annihilations.

\section{Gamma ray -- Dark Matter connection: FAQ}
\label{sec:faq}

The introductory material contained in this section is aimed at 
non-experts, and reviews the fundamental motivations for indirect
searches. 
The experienced reader can skip this section
and move 
directly to the next one, where we present an update on indirect searches. 

As mentioned in the introduction, a number of good reviews on
DM candidates and searches has recently appeared in literature. 
In the limited space available here, we will not attempt to provide 
an exhaustive review of the (many!) candidates proposed in 
literature, nor a description of the various detection strategies
proposed over the last two decades. Instead, it is probably more 
appropriate to take a step back, and tackle here some fundamental 
questions, frequently asked at conferences and workshops by 
non-experts, e.g.: if DM is {\it dark} by definition, 
why should one expect any indirect signature, such as emission of
gamma-rays, or neutrinos? Why most studies focus on annihilation
rather than, say, decay of DM particles? Why so many experiments 
exploring the GeV--TeV energy range, what motivates this mass scale? 
Finally, how {\it natural} are these scenarios? As we shall see, 
for some of these questions there is a precise answer, while for others the 
answer can only be tentative, reflecting our ignorance on what DM 
really is.   

\paragraph{{\bf Why annihilations?}}
DM is {\it dark} in the sense that there is no apparent electromagnetic
emission associated with it. So why should
one expect to detect any emission such as gamma-rays and neutrinos from DM?
Why every indirect DM search focuses on DM {\it annihilation} rather than, 
say, decay, or interactions with ordinary matter?

To answer this question, we start from the standard DM paradigm, 
commonly adopted by cosmologists, where DM is made of some new particles, 
{\it thermally produced} in the Early Universe,  kept in thermal 
and chemical equlibrium in the early Universe through pair annihilation 
into Standard Model particles, and Standard model processes leading to the
production of DM particle pairs. The density of DM particles $n(t)$
is governed  by the Boltzmann equation~\cite{Kolb:vq}
\begin{equation}
\frac{\mbox{d} n}{\mbox{d} t} + 3 H n= - < \sigma v > (n^2-n_{eq}^2)
\end{equation}
where H is the Hubble parameter, $<\sigma v>$ is the thermally averaged 
annihilation cross section, and $n_{eq}$ is the number density at thermal 
equilibrium. For particles of mass m in the non-relativistic limit,
$ n_{eq}\propto e^{-m/T}$. 

In practice, the density of  DM particles falls exponentially with the
temperature  until DM {\it freezes-out}, i.e. until DM falls out of
thermal equilibrium. This happens when the annihilation rate $\Gamma=<\sigma v> n$
drops below the expansion rate $H$. After freeze-out, the DM density
remains constant, and its value is usually expressed as $\Omega = \rho_{DM}/\rho_c$,
i.e. in units of the {\it critical density} $\rho_c= 3H^2/8\pi G$, where G 
is the Newton constant.

A useful approximation for the relic density of a DM particle with 
annihilation cross-section $<\sigma v>$ is
\begin{equation}
\Omega_{DM}=\frac{3 \times 10^{-27} \mbox{cm}^3 \mbox{s}^{-1}}{<\sigma v>}
\end{equation}
The appropriate relic density can thus be achieved with cross sections
typical of weak interactions. That's incidentally why one usual refers to 
DM candidates as WIMPs, fow Weakly Interacting Massive Particles. The 
calcuation of the actual relic density is usually complicated by particle
physics processes including, but not limited to, {\it co-annihilations}. 
We refer the interested reader to the review articles cited above for
more information and references.

In this framework, the fact that DM is {\it dark} in the local Universe, 
is due to the fact that it is far too diluted, on average, to produce 
observable annihilation product. However, this is true only {\it on average},
while the theory of structure formation, supported by N-body simulations,
predicts the existence of strong inhomogeneities, due to the gravitational
growth of density perturbations. 
Since the annihilation rate depends quadratically on the DM density, it is  
natural to search for DM annihilation products (photons, neutrinos etc)
by looking at regions where DM is expected to accumulate, reaching 
high densities, thus high annihilation rates, such as the center of 
galaxies, DM {\it clumps} and other targets that will be discussed below. 

We stress here that the WIMP paradigm, although appealing and theoretically
well motivated, is still tentative, and not conformed by any experimental
evidence. DM could be made of particles that never were in thermal equilibrium 
in the early universe, and in general it could exhibit a phenomenology
very different from the one described above. Indirect searches aim precisely 
at obtaining some insights on the nature of DM, to discriminate WIMP-like
scenarios from the many others proposed in literature.

\paragraph{{\bf Why gamma-rays?}}
The energy scale of the annihilation products is determined by the mass 
of the DM particles, as they typically carry a relatively large fraction,
say $O(0.1)$, of the available annihilation energy.  
DM candidates are commonly believed to have masses in the range
\begin{center}
\[ \;\;\;\;\;\;\;\;\;\;\;\;\;\;\;\;\; \;\;\;\;\;20 \mbox{GeV} \lesssim m_\chi \lesssim 120 \mbox{TeV} \]
\end{center}
The lower value corresponds to the so-called Lee-Weinberg 
limit \cite{leeweinberg,Hut:1977zn}, valid for fermionic candidates for 
which the annihilation cross section is proportional to their mass squared
$m_\chi^2$, which has been here updated using the current constraints
on the DM relic density. Basically, the lower the mass, the lower $\sigma v$, the higher the
relic density. Fermionic thermal relics lighter than $\approx 20$ GeV would overclose
the Universe, leading to unacceptable values of $\Omega_{DM}$. For
realistic, and theoretically well motivated, scenarios, a stronger lower
limit comes from null accelerator searches. In the case of 
Supersymmetric theories, the lower limit on the mass of the
neutralino (by far the most studied DM candidate), is around $\approx 30$ GeV,
the exact value depending on the specific supersymmetric
scenario adopted. Precision electroweak data
constrain the inverse compactification scale of Universal Exta Dimension models,
thus the mass of the LKP, 
to be larger than $\approx 300$ GeV~\cite{Gogoladze:2006br}.

The upper value comes from the so-called {\it unitarity 
bound} \cite{weinberg,Griest:1989wd}, which is here re-evaluated by
inserting in the old derivation the most recent estimates of the DM relic density. 

We stress that both limits are actually model dependent, and should be taken
with a grain of salt. Light particles, with O(MeV) mass,  have for instance been
proposed as viable DM candidates, evading the Lee-Weinberg
limit thanks to the {\it scalar} nature of these candidates (see below for more 
comments and a list of references). Other,
very massive, particles could violate the unitarity bound, an example
of these candidates are the so-called {\it wimpzillas}~\cite{Kolb:1998ki,Chang:1996vw}.

\paragraph{{\bf Why now?}} Indirect DM searches have been 
proposed over 25 years ago (an incomplete list of references includes 
Refs.~\cite{Gunn:1978gr,Stecker:1978du,Srednicki:1985sf,Rudaz:1986db,Bergstrom:1988fp,Stecker:1987dz,Stecker:1988ar,Bouquet:1989sr,Rudaz:1990rt}).
Since gamma-ray experiments such as EGRET didn't provide any 
conclusive evidence for DM
particles, why should we trust that present or future telescopes 
such as CANGAROO~\cite{cangaroo}, GLAST~\cite{glast}, HESS~\cite{hess}, 
MAGIC~\cite{magic} or VERITAS~\cite{veritas} will be more succesful? 
This question is related to another delicate
issue: what fraction of the observation time of a gamma-ray 
telescope should be devoted to DM searches, given the enormous
uncertainties associated to the predicted fluxes? The best answer
we can give here is that we learned a lot about DM in the last 
2 decades, especially on what DM is {\it not}. As we argue below, there
are indeed good reasons to think that the upcoming generation
of gamma-ray, neutrino and anti-matter telescopes, with their
capability of exploring new windows in the energy spectrum, 
and already explored windws with unprecedented sensitivity, may 
eventually find the smoking-gun for DM annihilations, a result 
that would  be of paramount importance for our understanding of
the Universe.

\section{Status Quo and Conflicting claims}
\label{sec:claims}

The difficulty of obtaining from astrophysical
observations conclusive answers on the nature of DM, is 
witnessed by the numerous conflicting claims of 
discovery, recently appeared in literature. A number of 
observations have been in fact ``interpreted'' in terms of
DM, without providing, though, conclusive enough evidence
to claim ``discovery''. 

\paragraph{{\bf MeV Dark Matter}} The first example of 
a ``conflicting claim'' concerns the so-called {\it
Light Dark Matter} scenarios. As stressed above, the 
Lee-Weinberg limit on the mass of the DM particle can be evaded 
if one e.g. postulates that the DM particle is a scalar. 
However, there is in principle no reason why one should
prefer such a candidate over the more famous, and theoretically
well motivated, neutralino. The situation changed, however, after 
the launch of the INTEGRAL satellite, due to the observation
of an intense 511 keV annihilation line from a region of
size $\approx 8 ^\circ$ centered around the galactic 
center~\cite{Knodlseder:2005yq}.
This emission did not come as a surprise 
to astrophysicists, who had discovered the electron-positron
annihilation feature already in the early seventies, making 
it the first extra-solar system spectral feature ever observed~\cite{johnson}. 
The INTEGRAL data however have reopened the old debate on
the origin of the positrons annihilating on ambient electrons.

Many astrophysical explanations have been proposed, including
production by black holes and pulsars~\cite{dermer88},
microquasars~\cite{Li:1995ev}, radioactive nuclei from past
supernovae, novae, red giants or Wolf-Rayet
stars~\cite{Casse:2003fh,signore88}, a single recent gamma-ray burst 
event~\cite{Furlanetto:2002sb}, cosmic ray interactions with the
interstellar medium~\cite{kozlowsky87} and stellar
flares~\cite{meynet97}. More recently, new scenarios have been
proposed invoking pulsar winds~\cite{Wang:2005cq}, 
primordial and accreting small-mass black
holes~\cite{Frampton:2005fk,Titarchuk:2005uw} and gamma-ray bursts~\cite{grbpaper}.

The large uncertainties associated with each of these scenarios,
however, left the door open to more
``exotic'' explanations. In particular, the possibility to explain the data 
in terms of from DM annihilation immediately 
attracted the attention of particle astrophysicists, also
in view of the lack of a disk component of the 511 keV emission,
a natural circumstance for a scenario where the positrons 
would  be emitted in a spherically symmetric fashion, following the
distribution of DM. However, a simple calculation suggested that
any candidate with a mass above the pion mass would inevitably 
produce gamma-rays and synchrotron emission far above the 
experimental data. In particular, if the 511 keV emission was
due to positrons produced by annihilation of neutralinos, 
the associated gamma-ray flux would exceed the observed EGRET 
flux by seven orders of magnitude! A Light DM candidate 
was instead shown to succesfully reproduce the normalization
of the observed 511 keV line without violating any other 
observational constraint~\cite{Boehm:2003bt}. Following this claim, 
the interpretation in terms of the annihilation or decay of
many other ``exotic'' candidates have been proposed in literature,
including
axinos~\cite{Hooper:2004qf}, sterile
neutrinos~\cite{Picciotto:2004rp}, scalars with gravitational strength
interactions~\cite{Picciotto:2004rp}, mirror
matter~\cite{Foot:2004kd}, color superconducting dark
matter~\cite{Oaknin:2004mn}, superconducting cosmic
strings~\cite{Ferrer:2005xv}, moduli ~\cite{Kawasaki:2005xj} and
Q-balls~\cite{Kasuya:2005ay}.

How to prove that this explanation was right? Clearly, the 
Light DM intepretation is to be considered {\it tentative} until
one can find a smoking-gun for it, or make a testable prediction.
The first prediction, i.e. the detection of an annihilation line
from a dwarf galaxy, has so far failed~\cite{Cordier:2004hf}
, while further analysises
have progressively reduced the allowed parameter space of DM 
particles. On one side, an upper limit on the mass comes from
the analysis of Internal Bremsstrahlung emission ($\approx 20$ MeV, see 
Ref.~\cite{BBB05}) and in-flight annihilation (of order $3-7$ MeV, see Refs. 
~\cite{BY06,Sizun:2006uh}). 
On the other side, an analysis based on the explosion of the
supernova SN1987A sets a lower limit of $\approx 10$ MeV,
thus apparently ruling out Light DM as a viable explanation 
of the 511 keV line~\cite{Fayet:2006sa}, at least in its 
most simple realization.
Recently, the constraint on Internal-Brehmsstralung has been 
challenged, and the claim has been made that it is still possible
to accomodate all existing constraints while still providing
a satisfactory explanation of the INTEGRAL data~\cite{Boehm:2006df}:
the debate is thus still open. Even if all constraints are evaded, 
however, some fundamental questions remain: how can one prove the ``exotic''
origin of the positrons, how can one discriminate among different 
candidates, how to convince a particle physicist that we are 
dealing with new physics? It appears clear that to promote the 
Light DM scenario 
from ``tentative intepretation'' to ``discovery''
additional evidence is needed, such as peculiar spectral 
features (e.g. a 2$\gamma$ line~\cite{Boehm:2006gu}), or 
discovery in collider searches.

\paragraph{{\bf The GeV Excess}} Another claim of (tentative) 
discovery has recently been made, based on the analysis of gamma-ray
data obtained by EGRET. The evidence in this case would be for WIMPs
with a mass of tens of GeV, producing through their annihilation
a ``bump'' in the Galactic gamma-ray emission around 1 GeV~\cite{deBoer:2006tv}. 
Although in principle very exciting, the emission is characterized by
a distribution which is
very different from the one na\"ively predicted by numerical simulations
(more intense towards the galactic center), being in the shape
of a ring around the galactic center. This is not sufficent of course to rule 
out this scenario, but there are still numerous difficulties
associated with this intepretation, that have been recently 
highlighted in Ref.~\cite{Bergstrom:2006tk}, in particular regarding the required
ring-shaped distribution of DM, as well as the apparent 
incompatibility with anti-proton measurements.  
As in the case of MeV DM, this doesn't mean that
the proposed interpretation is wrong, but simply that a 
different approach is needed to obtain conclusive evidence.

\paragraph{{\bf Draco}} Some preliminary results of the CACTUS
collaboration~\cite{CACTUS} have recently been interpreted as a possible evidence
for a DM annihilation signal from the Draco dwarf galaxy. Because
of the reduced amount of baryons in such a small astrophysical 
system, an exotic origin of the observed O(100) GeV emission was
certainly worth being explored. Several authors have investigated
this possibility
and concluded that the tentative detection was at odds with the 
expected annihilation signal, as well hard to reconcile with earlier
EGRET observations~\cite{Bergstrom:2005qk,Profumo:2006hs}. 
More recently, a re-analysis of the CACTUS data has shown no
evidence for an excess of gamma-rays above 100 GeV~\cite{mani}.

\paragraph{{\bf The Galactic center}} The last example of 
``conflicting claims'' is provided by the interpretation of
the gamma-ray source(s) coincident with the Galactic center,
in terms of DM annihilation. The discovery of an EGRET source
in the direction of Sgr A* was in fact a potentially perfect 
signature of the existence of particle DM, as thoroughly discussed
in Refs.~\cite{Stecker:88,Bouquet:89,Berezinsky:94,Bergstrom:97,Bertone:2001jv,Cesarini:2003nr,Fornengo:2004kj}. However, it was subsequently
realized that the EGRET source could have been slightly offset
with respect to the position of Sgr A*, a circumstance clearly
at odds with a DM intepretation~\cite{Hooper:2002ru}.
\begin{figure}
\centering
  \includegraphics[width=0.45\textwidth]{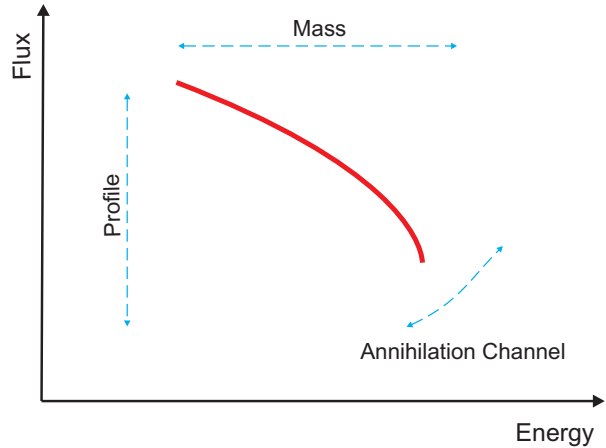}
\caption{The problem with indirect searches: the lack of constraints
on the mass scale, the profile and the leading annihilation
channel, leads to uncertainties on the energy scale and 
on the spectrum normalization and shape respectively.}
\label{fig:1}       
\end{figure}

Recently the gamma-ray telescope HESS has detected a high energy 
source, spatially coincident within $1'$ with Sgr A*~\cite{Aharonian:2004wa}
and with a spectrum extending above 20 TeV. 
Although the spatial coincidence is much more satisfactory 
than in the case of the EGRET source, the ``exotic'' origin
of the signal is hard to defend, since the implied mass scale 
of the DM particle (well above 20 TeV, to be consistent with the
observed spectrum) appears to be difficult to reconcile with 
the properties of commonly studied candidates , and the fact that 
the spectrum is a power-law, then, points towards a standard
astrophysical source (see e.g. the discussion
Ref.~\cite{Profumo:2005xd}). The galactic center,
however, remains an interesting target for GLAST, since it 
will explore a range of energies below the relatively high 
threshold of HESS, where a DM signal could be 
hiding~\cite{Zaharijas:2006qb}. The recent claim that the profile of 
large galaxies could be much more shallow than previously 
thought~\cite{Mashchenko:2006dm}, should not discourage
further studies, especially in view of the possible enhancement
of the DM density due to interactions with the stellar 
cusp observed at the Galactic center~\cite{MHB}. 

The detection of a signal from the Galactic center would be
extremely interesting, but can it prove the existence of DM? 
Realistically, one may hope to observe, at most, a ``bump'' above the
background. Without peculiar spectral features it would be hard 
to claim discovery of DM, unless a fit of the spectrum points
towards a mass compatible with the eventual findings of 
new physics searches at accelerators. 

Figure 1 illustrates the 
difficulties associated with the unambiguous identification
of a DM signal. Any excess, at any energy, could in principle 
be explained in terms of DM particles with appropriate properties:
the normalization of the flux can be adjusted by changing the
distribution of DM particles, the energy scale can be varied
over several orders of magnitude, taking advantage of our 
ignorance on the DM mass scale; even the slope can be modified,
since different annihilation channels lead to different spectra.

This doesn't mean that the tentative identifications presented
above are ruled-out: the signature of DM could have been already found in 
one or several sets of data, and all the above claims should be 
taken seriously and further investigated without prejudice, 
especially in view of the fact that we don't know what 
DM is! However, it is important to look for clear smoking-gun 
of DM annihilation, and study theoretical scenarios with 
unambiguous signatures that can be tested 
with present and future experiments. To this aim, we summarize in the next section
some recently proposed ideas that go precisely in this direction,
and that may shed new light on the nature of particle DM.

\begin{figure}
\centering
  \includegraphics[width=0.45\textwidth]{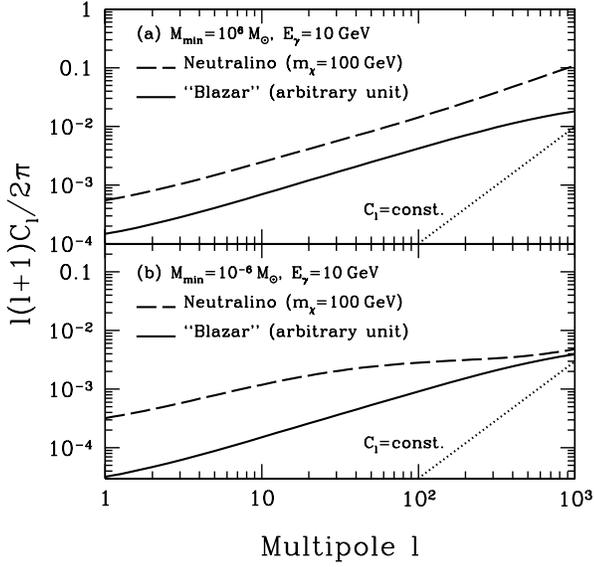}
\caption{Shape of the angular power spectrum of the CGB expected 
 from unresolved blazar-like sources
 (solid lines)
 with arbitrary normalizations.
 The power spectrum from annihilation of neutralinos 
 with $m_\chi = 100$ GeV is also plotted as the dashed lines. 
 The adopted gamma-ray energy is
 10 GeV, and the minimum mass of DM halo 
 is (a) $10^6 M_\odot$, and (b) $10^{-6} M_\odot$. 
 The dotted lines show the shot 
 noise ($C_l={\rm const.}$) with arbitrary  normalizations,
 which represent the power spectrum of very rare sources.
 From Ref.~\cite{Ando:2005xg}}
\label{fig:komatsu}       
\end{figure}

\section{New strategies}
\label{sec:smoking}

Before starting the discussion of new strategies for the
 unambiguous detection of DM, we recall the first, and  more
clear signature that one may hope to detect: distinctive
spectral features, and in particular annihilation lines. 
This has been discussed thoroughly in literature, and 
although it appears unlikely that commonly discussed 
candidates such as the supersymmetric neutralino, possess
prominent enough feature to be detected with current or
upcoming experiments, it is probably good to keep this 
possibility in mind, and to search future gamma-data 
for signatures of this kind.
\begin{figure}[t]
\centering
  \includegraphics[width=0.45\textwidth]{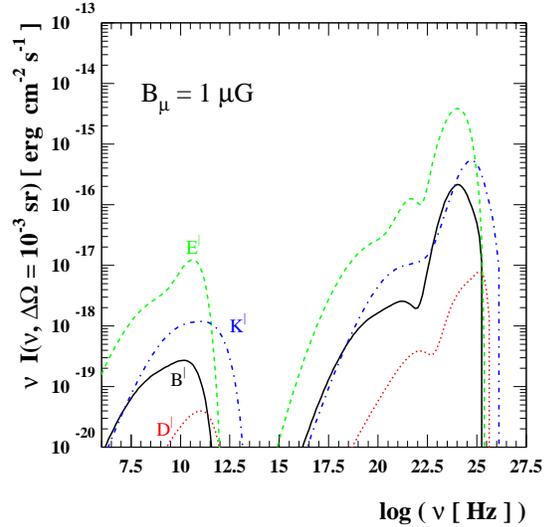}
\caption{Multi-wavelength spectra for four different benchmark DM models,
for a best fit NFW profile, and a mean magnetic field equal to $1 \mu G$. 
From Ref.~\cite{cola1}, see {\it ibid.} for more details.}
\label{fig:cola}       
\end{figure}
\subsection{Gamma-ray background}

Although most searches have focused on the identification of
point-sources associated with regions where DM accumulates,
it is interesting to ask what the gamma-ray background produced
by the annihilations of DM in all structures, at any redshift, 
would be. The first calculation of this type was performed 
in Ref.~\cite{Bergstrom:2001jj} , and then further studied in Refs.~\cite{Taylor:2002zd,Ullio:2002pj}. 
The annihilation background can be expressed as
\begin{equation}
\Phi(E)= \frac{\Omega_{DM}^2 \rho_c^2}{8 \pi H_0}  \frac{\sigma v}{m_\chi^2}
\int_0^{z_{max}} \mbox{d}z \frac{\Delta^2}{h(z)} N(E')
\end{equation}
where $N(E')$ is the gamma-ray spectrum per annihilation,
$H_0$ is the Hubble parameter and $h(z)=[(1+z)^3\Omega_{DM}+\Omega_\Lambda]^{1/2}$.
The information on the shape of individual DM halos in 
encoded in $\Delta^2$, which is essentially the integral 
of $\rho^2$ over the virial volume of the halo. 
Although it is unlikely that the annihilation
background will be detected without first detecting a prominent 
gamma-ray source at the Galactic center~\cite{Ando:2005hr}, the
characteristic power spectrum of the gamma-ray background
would discriminate its DM origin from ordinary astrophysical
sources~\cite{Ando:2005xg}. 
\begin{figure*}[t]
\centering
  \includegraphics[width=0.75\textwidth]{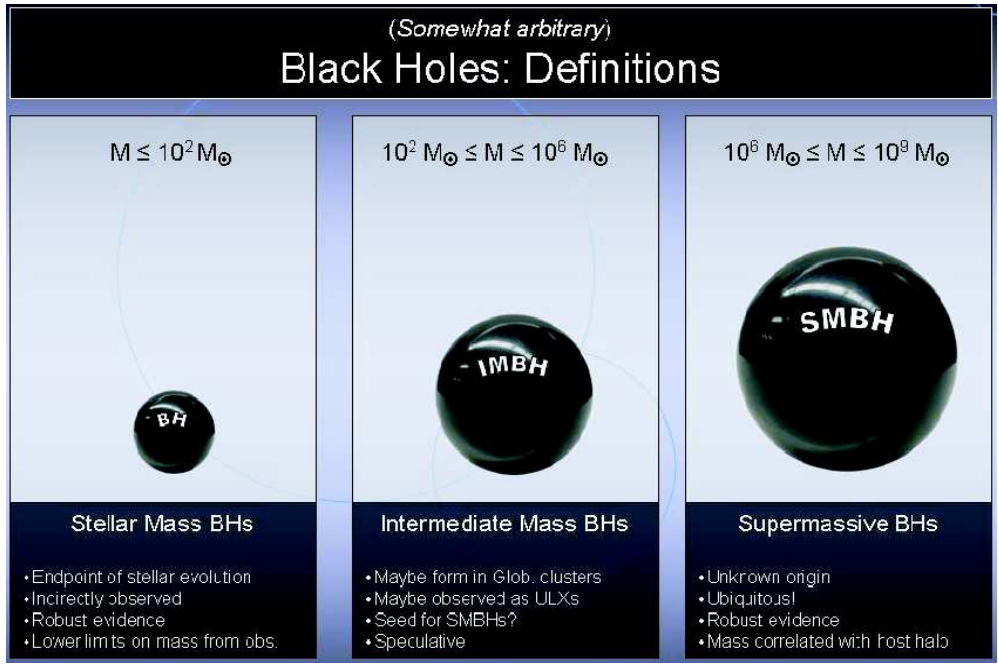}
\caption{Classification of Black Holes by mass. There is robust 
evidence for Stellar mass and Supermassive Black Holes, while  
Intermediate mass Black Holes are still speculative. IMBHs and SMBHs
may act as ``annihilation boosters'', with important
implications for DM searches.}
\label{fig:bhs}       
\end{figure*}
We show in fig.~\ref{fig:komatsu}   
 the power spectrum of the gamma-ray background produced 
by annihilation of neutralinos 
 with $m_\chi = 100$ GeV, compared with the one relative to 
 unresolved blazar-like sources.
Above $l\sim 200$ the DM spectrum continues to grow whereas 
the blazar spectrum flattens out,
due to the cut-off adopted by the authors corresponding to
 the minimum mass of halos hosting blazars ($\approx  10^{11} M_\odot$).
The annihilation spectrum thus appear to have much more power
at large angular scales, which should be easily distinguished
from the blazar spectrum.

There are large uncertainties associated with
this calculation, mainly due to our ignorance of the DM profile 
in the innermost regions of halos, and of the amount of substructures.
The existence of mini-spikes (see below) would also dramatically affect 
the predicted result~\cite{Horiuchi:2006de}. But the clear prediction is
made that if the observed background has the peculiar shape discussed
above, this may be consider as a hint of DM annihilations.
Recently the calculation of the {\it neutrino} background from
DM annihilations has been performed, adopting a formalism very similar
to the one sketched above. The comparison with observational data 
allows to set an interesting, and very general, 
upper bound on the DM total annihilation cross section~\cite{Beacom:2006tt}.

\subsection{Multi-wavelength}

An alternative strategy is to employ a multi-messenger, multi-wavelength
approach. In fact, despite the freedom in the choice of DM
parameters makes the intepretation of observational data rather
inconclusive, one can always combine the information at different
wavelengths, and with different messengers, to obtain more
stringent contraints. In fact, gamma-rays are typically (but not
exclusively) produced
through annihilation and decay chain involving neutral pions
\begin{equation}
\chi \overline{\chi} \rightarrow q \overline{q} \rightarrow \left[ \mbox{fragmentation} \right] \rightarrow \pi^0 \rightarrow 2\gamma
\end{equation}
Every time gamma-rays are produced this way, leptons 
and neutrinos are also produced folloeing the chain
\begin{equation}
\chi \overline{\chi} \rightarrow q \overline{q} \rightarrow \mbox{fragmentation} \rightarrow \pi^\pm \rightarrow l, \nu_l, ... 
\end{equation}
An example of this approach is the combined
study of the gamma-ray emission from the Galactic center and 
the associated synchrotron emission produced 
by the propagation of electron-positron pairs in the Galactic
magnetic field~\cite{Gondolo:2000pn,Bertone:2001jv,Bertone:2002ms,Aloisio:2004hy}. Similarly one can investigate what the flux of 
neutrinos would be, once the gamma-ray flux has been normalized to
the EGRET data~\cite{Bertone:2004ag}. 

One can also ask what the fate 
of the electron-positron pairs produced by DM annihilation is in 
dwarf galaxies and clusters of galaxies. An example of this 
approach can be found in Refs.~\cite{cola1,cola2}, where the authors study
the synchrotron and gamma-ray emission from Draco and from 
the Coma cluster. In fig.~\ref{fig:cola} we show 
the multi-wavelength spectra of Draco, relative to four diffferent 
DM benchmark models, assuming a NFW profile and a mean
magnetic field of $1 \mu G$.

\begin{figure*}[ht]
\centering
  \includegraphics[width=0.75\textwidth]{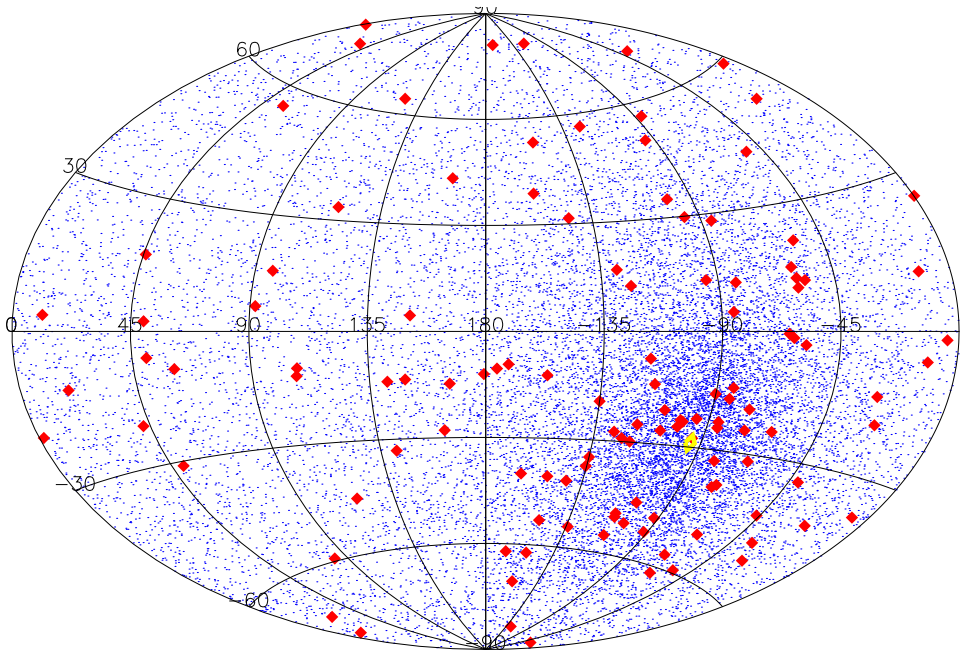}
\caption{Sky map in equatorial coordinates showing the position 
of Intermediate Mass Black Holes in one random realization
of a Milky-Way like halo (red diamonds), and in all 200 realizations (blue dots).
The concentration at negative declinations corresponds to 
the position of the Galactic center (black open diamond).
From Ref.~\cite{Bertone:2006nq}}
\label{fig:sky}       
\end{figure*}
A word of caution is in order, however, when combining information
relative to different wavelengths. In fact, not only the available
data, due to the different angular resolution of experiments, are
relative to different physical regions. But the calculation of
the associated spectra at different energies usually requires 
further inputs, thus introducing new parameters to the problem.
The aforementioned calculation of the synchrotron emission is a 
typical example: Although for every specific DM model, the
 number of electron-positron pairs produced per annihilation is
fixed, the calculation of the synchrotron emission requires an estimate 
of the diffusion of postrons and it further depends on the magnetic
field profile, typically poorly constrained on the scales of interest.

\subsection{Mini-spikes}

Among the new strategies discussed in this section, we will devote 
particular attention to the effect of black holes (BHs) growth on the
surrounding distrbution of DM, a circumstance that will lead us to conclude that
BHs can act as DM annihilation ``boosters''. 

Black Holes (BHs) can be broadly divided in 3 different classes, as 
schematically illustrated in fig.~\ref{fig:bhs}. The first
class include BHs with mass smaller than $\approx 100$
solar masses, tpypically remnants of the collapse of 
massive stars (recent simulations suggest that the upper 
limit on the mass of these objects is as low 
as $\approx 20 M_\odot$ ~\cite{Fryer:2001}).
There is robust evidence for the existence of these objects,
coming from the observation of binary objects with
compact objects whose mass exceeds the critical mass of 
Neutron Stars. For a review of the topic and the discussion of the
possible smoking-gun for Stellar Mass BHs see e.g.~\cite{Narayan:2003fy} 
and references therein.

The existence of Supermassive BHs (SMBHs) , lying at the center of
galaxies (including our own), is also well-established 
(see e.g. Ref.~\cite{Ferrarese:2005}),
and intriguing correlations are observed between the BHs mass
and the properties of their host galaxies and 
halos~\cite{kormendy:1995,Ferrarese:2000se,McLure:2001uf,Gebhardt:2000fk,Tremaine:2002js,Koushiappas:2003zn}. From a theoretical point of view, a population of massive seed 
black holes could help to explain the origin of SMBHs. 
In fact, observations of quasars at 
redshift $z\approx 6$ in the 
Sloan Digital survey ~\cite{Fan:2001ff,barth:2003,Willott:2003xf}
suggest that SMBHs were already in place 
when the Universe was only $\sim 1$ Gyr old, a circumstance 
that can be understood in terms of rapid growth starting 
from ``massive'' seeds (see e.g. Ref.~\cite{haiman:2001}).  

This leads us to the third category of BHs, characterized 
by their {\it intermediate} mass. In fact, scenarios 
that seek to explain the properties of the observed 
supermassive black holes population  result in the prediction
of a large population of wandering Intermediate Mass BHs (IMBHs). Here, following 
Ref.~\cite{Bertone:2005xz}, we consider two different formation
scenarios for IMBHs. In the first scenario,
IMBHs form in rare, 
overdense regions at high redshift, 
$z \sim 20$, as remnants of Population III stars, and have a 
characteristic mass-scale of a few $10^2 M_\odot$ 
\cite{Madau:2001} (a similar scenario
was investigated in Ref.~\cite{Zhao:2005zr,islamc:2004,islamb:2004}).  
In this scenario, these black holes serve as the 
seeds for the growth supermassive black 
holes found in galactic spheriods \cite{Ferrarese:2005}.  
In the second scenario, IMBHs form directly out of cold gas 
in early-forming halos 
and and are typified by a larger mass scale, 
of order $10^5 M_\odot$~\cite{Koushiappas:2003zn}. 
In Fig.~\ref{fig:sky} we show the distribution of IMBHs
in the latter scenario, as obtained in Ref.~\cite{Bertone:2006nq}.

We have so far discussed about BHs, but we haven't yet established the
connection with DM searches. The effect of the formation of a central 
object on the surrounding distribution of matter has been investigated 
in Refs.~\cite{peebles:1972,young:1980,Ipser:1987ru,Quinlan:1995} 
and for the first time in the framework of DM annihilations in 
Ref.~\cite{Gondolo:1999ef}. It was shown that  the 
{\it adiabatic} growth of a massive object at the center of a 
power-law distribution of DM with index $\gamma$, induces 
a redistribution of matter into a new power-law (dubbed ``spike'') with index
$\gamma_{sp}=(9-2\gamma)/(4-\gamma)$
This formula is valid over a region of size $R_s \approx 0.2 r_{BH}$, 
where $r_{BH}$ is the radius of gravitational influence
of the black hole, defined implicitly as $M(<r_{BH})=M_{BH}$,
with $M(<r)$ mass of the DM  distribution within a 
sphere of radius $r$, and $M_{BH}$ mass of the Black Hole
~\cite{Merritt:2003qc}.
The process adiabatic growth is in particular valid for the SMBH at the 
Galactic center. A critical assessment of the formation {\it and survival}
of the central spike, over cosmological timescales, is presented
in Refs.~\cite{Bertone:2005hw,Bertone:2005xv} (see also references therein).

Here we will not further discuss the spike at the Galactic center, 
and will rather focus our attention on {\it mini-spikes} around IMBHs.
If $N_{\gamma}(E)$ is 
the spectrum of gamma-rays per 
annihilation, the gamma-ray flux from an individual mini-spike 
can be expressed as~\cite{Bertone:2005xz}
\begin{equation}
\Phi_{\gamma}(E)  = \phi_0  m_{\chi,100}^{-2} (\sigma v)_{26} D_{\rm  kpc}^{-2} L_{\rm sp} N_{\gamma}(E)
\label{eq:intrinsic}
\end{equation}
with $\phi_0 = 9 \times 10^{-10} {\rm cm}^{-2}{\rm s}^{-1}$.  
The first two factors depend on the particle physics parameters,
viz. the mass of the DM particle in units of 100 GeV 
$m_{\chi,100}$, and its annihilation
cross section in units of $10^{-26} {\rm cm}^3/{\rm s}$, $(\sigma v)_{26}$,
while the third factor accounts for the flux dilution with 
the square of the IMBH distance to the Earth in kpc, $D_{\rm  kpc}$. 
Finally, the normalization of the flux is fixed by an adimensional 
{\it luminosity factor} $L_{\rm sp}$, that depends on the specific
properties of individual spikes. In the case where the DM profile 
{\it before} the formation of the IMBH follows the commonly
adopted Navarro, Frenk and White profile~\cite{Navarro:1996he},
the final DM density $\rho(r)$ around the IMBH will be described by a 
power law $r^{-7/3}$ in a region of size $R_s$
around the IMBHs. Annihilations themselves will set an upper limit
to the DM density $\rho_{\rm max}\approx m_\chi/[(\sigma v) t]$, where 
$t$ is the time elapsed since the formation of the mini-spike, and
we denote with $R_{\rm c}$ the ``cut'' radius where $\rho(R_{\rm c})=\rho_{\rm max}$. 
With these definitions, the intrinsic luminosity factor  
in Eq.~\ref{eq:intrinsic} reads
\begin{equation}
L_{\rm sp}\equiv
\rho^{2}_{100}(R_{{\rm s}}) R_{\rm s,pc}^{14/3}
R^{-5/3}_{\rm c,mpc}
\end{equation}
where $R_{{\rm s,pc}}$ and $R_{\rm c,mpc}$ denote respectively
 $R_{{\rm s}}$ in parsecs and $R_{\rm c}$ in units of $10^{-3}$pc,
$\rho_{100}(r)$ is the density in units of $100$GeV cm$^{-3}$.
Typical values of $L_{\rm sp}$ lie in the range 0.1 -- 10~\cite{Bertone:2005xz}.

\begin{figure}
\centering
  \includegraphics[width=0.45\textwidth]{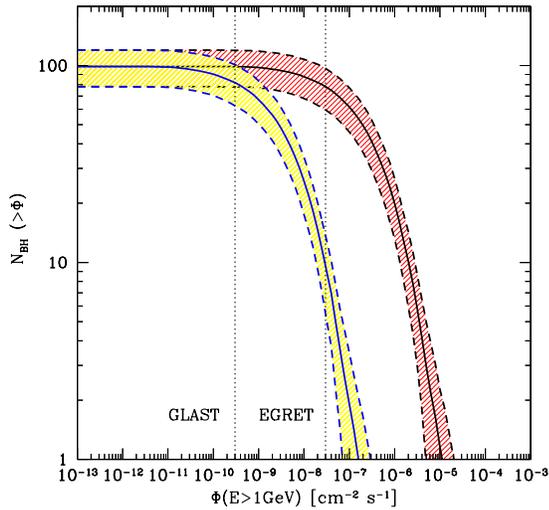}
\caption{IMBHs integrated luminosity function,
(number of mini-spikes detectable with an experiment of sensitivity
$\phi$) for IMBHs with mass $\sim 10^5 M_\odot$.
The upper (lower) line corresponds to $m_\chi=100$ GeV, $\sigma v=3\times 10
^{-26}$ cm$^3$ s$^{-1}$ ($m_\chi=1$ TeV, $\sigma v= 10
^{-29}$ cm$^3$ s$^{-1}$). For each curve we also show the 1-$\sigma$
scatter among different realizations of Milky Way-sized host DM halos.  
We show for comparison the 5$\sigma$ point 
source sensitivity above $1$~GeV of EGRET and GLAST (1 year). From Ref.~\cite{Bertone:2005xz}}
\label{fig:detected}       
\end{figure}

In Fig.~\ref{fig:detected}, we show the (average) integrated luminosity 
function of IMBHs in scenario B.  We define the integrated luminosity 
function as the number of black holes producing a gamma-ray 
flux larger than $\Phi$, as a function of $\Phi$. Loosely
speaking, this can be understood as he number of mini-spikes
that can be detected with an experiment with point source sensitivity $\Phi$ above 1~GeV.   
The upper (lower) line corresponds to $m_\chi=100$~GeV, 
$\sigma v=3\times 10^{-26}$~cm$^3$s$^{-1}$ 
( $m_\chi=1$~TeV, $\sigma v=10^{-29}$~cm$^3$s$^{-1}$).  
We show for comparison the point source sensitivity 
above 1~GeV for EGRET and GLAST, corresponding roughly 
to the flux for a $5\sigma$ detection of a high-latitude 
point-source in an observation time of 
1~year~\cite{Morselli:2002nw}.  
The dashed region
corresponds to the $1\sigma$ scatter between different 
realizations of Milky Way-sized halos.
This band 
includes the variation in spatial distributions of IMBHs 
from one halo to the next as well as the variation in 
the individual properties of each IMBH in each realization.
GLAST may thus be able to detect up to 100 point sources 
with identical spectra, and not correlated with the 
disk, thus providing compelling evidence for the 
non-astrophysical origin of their gamma-ray emission.
If DM is heavy, above $\approx 300$ GeV, Air Cherenkov Telescopes 
can be used to extend the observation of GLAST sources 
to higher energies. At the same time, the annihilation
of heavy particles to neutrinos may lead to interesting 
signatures in neutrino telescopes~\cite{Bertone:2006nq}.


\section{Conclusions}
To establish the connection between gamma-ray astrophysics and DM 
searches, we have first reviewed the fundamental motivations for indirect
DM searches. We tried to convey the fundamental message, especially to 
non-experts, that despite the ``exotic'' nature of the particles under
consideration, indirect signals of DM annihilation are a ``natural''
expectation in the standard WIMP paradigm, and that it is certainly worth 
pursuing further indirect DM searches. 
We have also provided a critical assessment of some recent ``conflicting 
claims'' of discovery, recently appeared in literature. Without
taking sides in the disputes over specific DM candidates, we have 
stressed the difficulty of making unambiguous claims, especially 
in view of the large uncertainties associated with the mass scale, 
the distribution and the nature of DM. We have finally shown that
despite these conflicting claims, the ``status quo'' shouldn't 
discourage further studies, especially
in view of the numerous new strategies that have been recently 
proposed, and that may provide unambiguous evidence for DM. Among
them we have discussed the peculiar power-spectrum
of the gamma-ray background from DM annihilations; the multi-wavelength,
multi-messenger approach; and the so-called ``mini-spike'' 
scenario, where IMBHs act as ``boosters'' of the DM annihilation signal.

\begin{acknowledgements}
I thank my collaborators
J. Beacom, N. Bell, D. Merritt, E. Nezri, J. Orloff, G. Servant, 
G. Sigl, J. Silk and  A. Zentner. It is a pleasure to thank the
organizers of the Barcelona Workshop for a stimulating and
well organized conference. This work was supported by the Helmholtz 
Association of National Research Centres, project VH-NG-006. 

\end{acknowledgements}



\end{document}